\documentstyle[epsfig,a4]{article}

\begin{document}

\begin{titlepage}

\centerline{\Large \bf Large-volatility dynamics in financial markets}

\vskip 2.0truecm
\centerline{\bf X.F. Jiang, B. Zheng\footnote{corresponding author; email: zheng@zimp.zju.edu.cn} and J. Shen}
\vskip 0.2truecm
\centerline{Physics Department, Zhejiang University, Hangzhou 310027, P.R. China}

\vskip 2.5truecm

\abstract{
We investigate the large-volatility dynamics in financial
markets, based on the minute-to-minute and daily data of the Chinese Indices
and German DAX. The dynamic relaxation both before and after large
volatilities is characterized by a power law, and the exponents
$p_\pm$ usually vary with the strength of the large volatilities.
The large-volatility dynamics is time-reversal symmetric at the
time scale in minutes, while asymmetric at the daily time scale.
Careful analysis reveals that the time-reversal asymmetry is mainly
induced by exogenous events. It is also the exogenous events which
drive the financial dynamics to a non-stationary state. Different characteristics of
the Chinese and German stock markets are uncovered.
}

\vspace{0.5cm}


\vspace{0.3cm}


\end{titlepage}

\section{Introduction}

Financial markets are complex systems which share common features
with those in traditional physics. In recent years, large amounts of high-frequency data have piled up in stock markets.
This allows an analysis of
the fine structure and interaction of the financial dynamics, and
many empirical results have been documented
\cite{man95,gop99,liu99,bou01,bou03,sor03,gab03,qiu06,she09,she09a}.
Although the price return of a financial index is short-range
correlated in time, the volatility exhibits a long-range temporal
correlation \cite{gop99,liu99}. The dynamic behavior of volatilities
is an important topic in econophysics \cite{gop99,liu99,yam05,wan06}.

Assuming that the financial market is in a stationary state,
one may analyze its static statistical properties. For a comprehensive understanding of the
financial markets, however, it is also important to investigate the
non-stationary dynamic properties. A typical example is the
so-called financial crash \cite{sor96,sor03}. Lillo and Mantegna
study three huge crashes of the stock market, and find that the rate
of volatilities larger than a given threshold after such market
crashes decreases by a power law with certain corrections in shorter
times \cite{lil03}. This dynamic behavior is analogous to the
classical Omori law, which describes the aftershocks following a
large earthquake \cite{omo94}. Selcuk analyzes the daily data of the
financial indices from 10 emerging stock markets and also observed
the Omori law after the two largest crashes \cite{sel04}. Mu and Zhou extend
such an analysis to the financial market in Shanghai \cite{mu08}. Recently,
Weber et al. demonstrate that the Omori law holds also after
"intermediate shocks", and the memory of volatilities is mainly
related to such relaxation processes \cite{web07}.

Stimulated by these works, we systematically analyze the
{\it large-volatility} dynamics in financial markets, based on the
minute-to-minute and daily data of the Chinese Indices and German DAX. In
our study, a large volatility is identified when it is sufficiently
large compared with the average one. Such an event maybe not yet corresponds to a real
financial crash {\it or rally}. In fact, the dynamic behavior of rallies has not been touched to our knowledge.
The purpose of this paper is multi-folds.
We investigate the dynamic relaxation both {\it before} and {\it
after} large volatilities. We focus on the time-reversal symmetry or
asymmetry at the different time scales. To
achieve more reliable results, we introduce the remanent and
anti-remanet volatilities to describe the large-volatility dynamics, different from those in Refs. \cite{lil03,sel04,mu08}.
In particular, we examine the dynamic behavior of different
categories of large volatilities, and search for the origin of the
time-reversal asymmetry at the daily time scale. We reveal how the
dynamic system is driven to a non-stationary state by exogenous
events. We compare the results of the mature German market and the
emerging Chinese market.

\section{Large-volatility dynamics}
In this paper, we have collected the daily data of the German DAX
from 1959 to 2009 with 12407 data points, and the minute-to-minute data from
1993 to 1997 with 360000 data points. The daily data of the Shanghai
Index are from 1990 to 2009 with 4482 data points, and the minute-to-minute
data are from 1998 to 2006 with 95856 data points. The daily data of
the Shenzhen Index are from 1991 to 2009 with 4435 data points, and
the minute-to-minute data are from 1998 to 2003 with 50064 data points. The
minute-to-minute data are recorded every minute in the German market, while
every 5 minutes in the Chinese market. A working day is about 450
minutes in Germany while exactly 240 minutes in China. In our
terminology, the results of the "Chinese Indices" are the
averages of the Shanghai Index and Shenzhen Index.

Denoting a financial index at time $t$ as $P(t)$, the return and
volatility are defines as $R(t)\equiv \ln P(t+1)-\ln P(t)$ and
$|R(t)|$ respectively. Naturally, the dynamic behavior of
volatilities may depend on the time scale. To study the dynamic
relaxation after and before large volatilities, we introduce the
remanent and anti-remanent volatilities $v_{+}(t)$ and $v_{-}(t)$,
\begin{equation}
v_{\pm}(t) = [< |R(t'\pm t)|>_c - \sigma]/Z
                  , \label{e10}
\end{equation}
where $Z=< |R(t')|>_c - \sigma$, $\sigma$ is the average volatility,
and $<\cdots>_c$ represents the average over those $t'$ with
specified large volatilities. In our analysis, the large
volatilities are selected by the condition $|R(t')|>\zeta$, and the
given threshold $\zeta$ is well above $\sigma$. When $\zeta$ is
sufficiently large, it corresponds to the financial crashes and
rallies. $v_+(t)$ describes how the system relaxes from a large
volatility to the stationary state, while $v_-(t)$ depicts how the
system approaches a large volatility.

Large shocks in volatilities are usually followed by a series of
aftershocks. Thus we assume that both $v_{+}(t)$ and $v_{-}(t)$ obey
a power law,
\begin{equation}
v_\pm(t) \sim (t+\tau_\pm)^{-p_\pm}, \label{e20}
\end{equation}
where $p_\pm$ are the exponents and $\tau_\pm$ are positive
constants. In most cases studied in this paper, the constants
$\tau_\pm$ are rather small. For reducing the fluctuations, we integrate Eq. (\ref
{e20}) from $0$ to $t$. Thus the cumulative function of $v_\pm(t)$
is written as
\begin{equation}
V_\pm(t) \sim [(t+\tau_\pm)^{1-p_\pm}-\tau_\pm^{1-p_\pm}]
\label{e30}
\end{equation}
for $p_\pm \neq 1$. The power-law behavior
of $v_{\pm}(t)$ just represents the long-range temporal
correlation of volatilities. Such a power-law behavior has been well
understood in dynamic critical phenomena, even in the case far from
equilibrium \cite {zhe98,zhe99}.

\begin{figure}[p]\centering
\epsfysize=7.cm
\epsfclipoff
\fboxsep=0pt
\setlength{\unitlength}{0.7cm}
\begin{picture}(9,18)(0,0)
\put(-2,12.50){{\epsffile{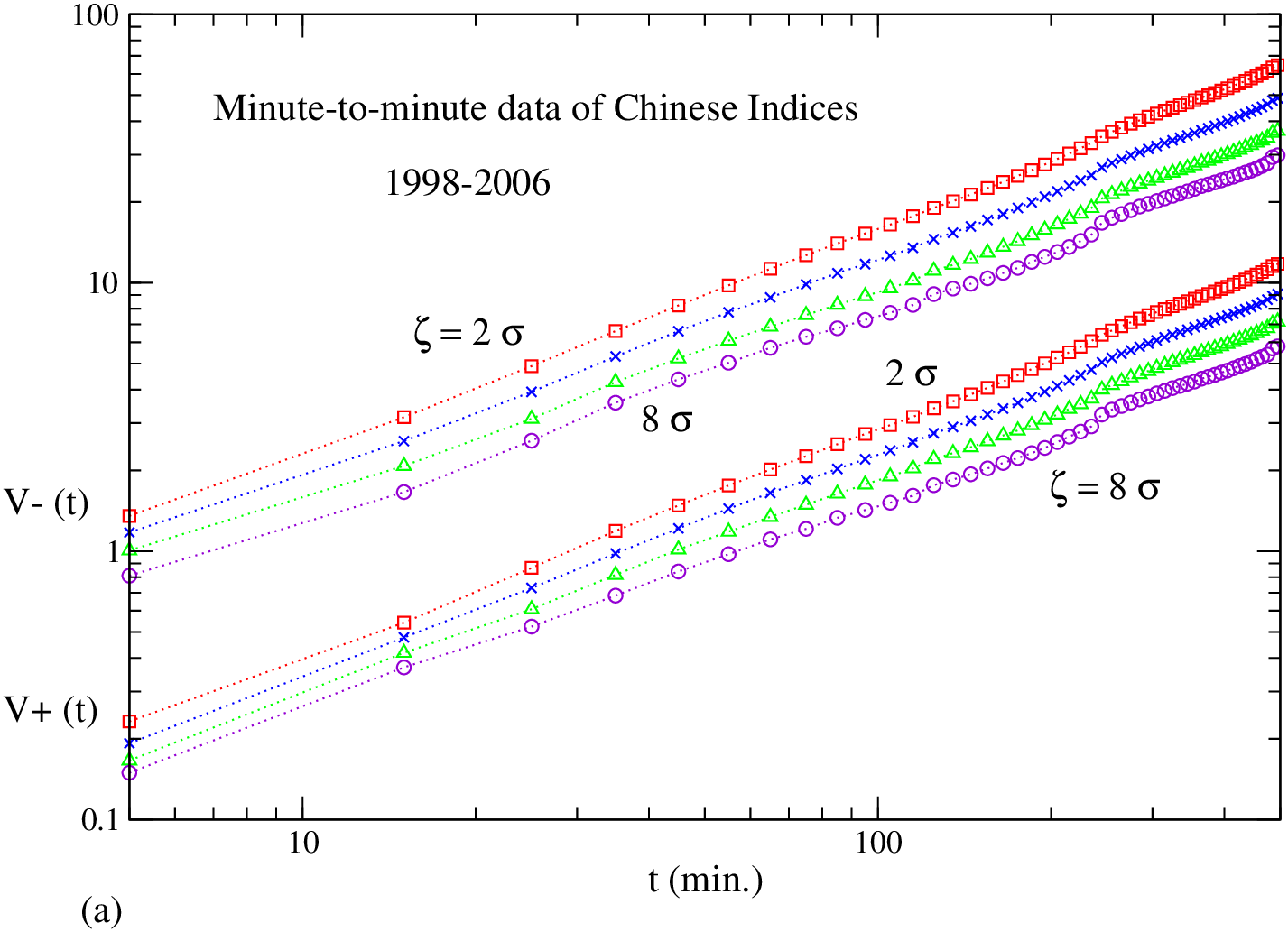}}}
\put(-2,0.50){{\epsffile{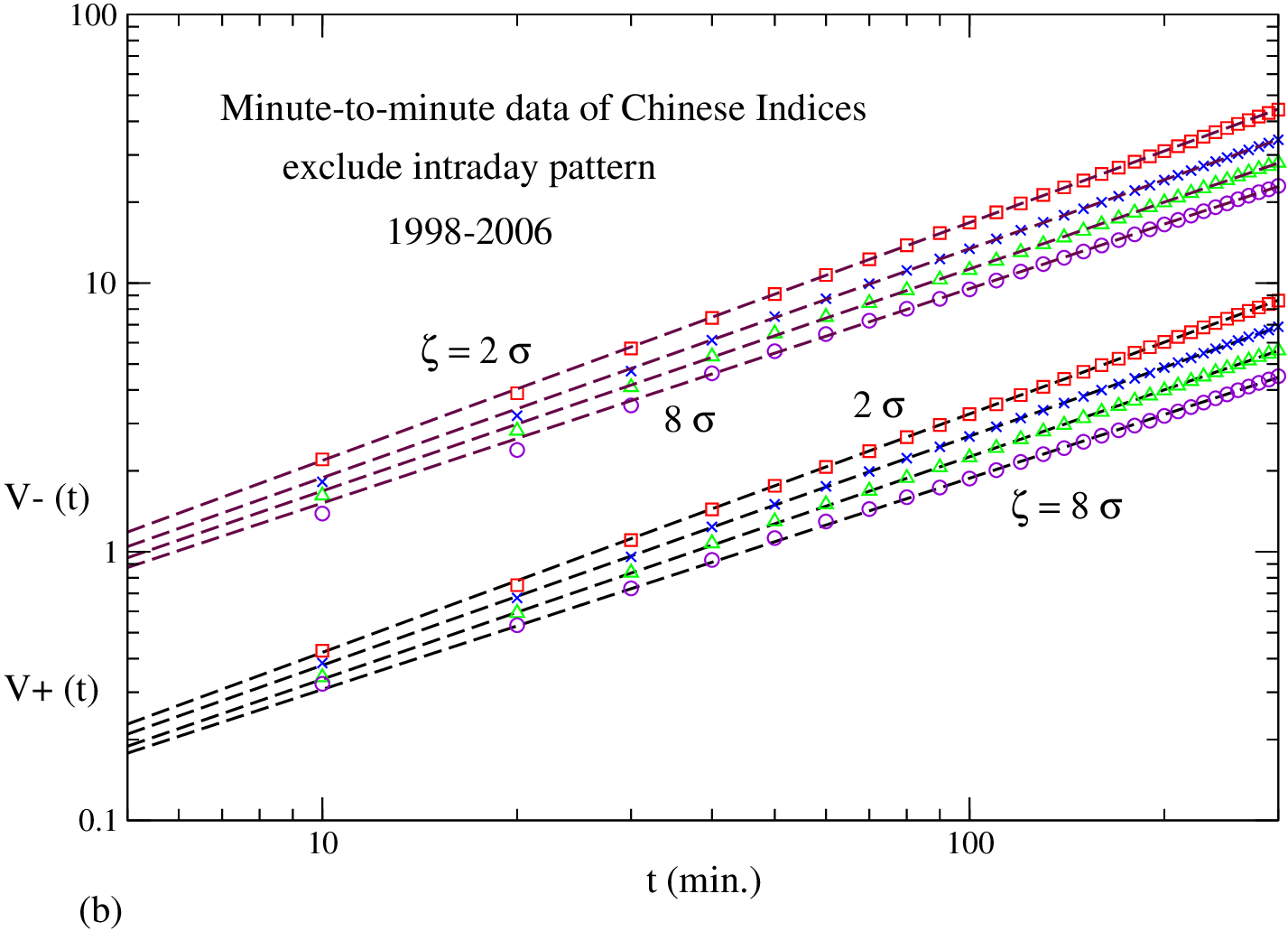}}}
\end{picture}
\caption{(a)
$V_\pm(t)$ for the minute-to-minute data of the Chinese Indices. From above,
the threshold is $\zeta= 2\sigma$, $4\sigma$, $6\sigma$ and
$8\sigma$ respectively. (b) The same as (a), but the intra-day
pattern has been removed. Dashed lines show the power-law fits with
Eq. (\ref {e30}). The $\zeta$-dependent exponents $p_-=p_+$, and
$\tau_\pm=0$.
}
\label{f1}
\end{figure}

We first analyze the minute-to-minute data of the Chinese Indices and German
DAX. Now $|R(t)|$ is calculated in the unit of five minutes for the
Chinese Indices and one minute for the German DAX. To select the
large volatilities, we set the threshold $\zeta=2\sigma$, $4\sigma$,
$6\sigma$, and $8\sigma$. For the minute-to-minute data, a large volatility
may not indicate a real macroscopic crash or rally, and it possibly
brings the dynamic system to a {\it microscopic} non-stationary
state. In Fig. \ref {f1} (a), $V_{\pm}(t)$ of the Chinese Indices
are displayed on a log-log scale. Because of the intra-day pattern
\cite{woo85,adm88,liu99,qiu06}, the curves periodically fluctuate at
a working day, i.e., $t\sim240$ minutes. We remove this intra-day
pattern, e.g., with the procedure in Ref. \cite{liu99,qiu07} and
recalculate $V_{\pm}(t)$. This is shown in Fig. \ref {f1} (b). Now
an almost perfect power-law behavior is observed for both $V_{-}(t)$
and $V_{+}(t)$, starting from $t \sim 5$ minutes. The curves of
$V_{\pm}(t)$ of the German DAX look very similar to those of the
Chinese Indices, with the intra-day pattern around $t\sim 450$
minutes.

Fitting the curves of $V_{\pm}(t)$ of the minute-to-minute data to
Eq.~(\ref {e30}), we obtain the exponents $p_{\pm}$ summarized in
the first and second sectors of Table~\ref {t1}. For both the
Chinese Indices and German DAX, both $p_{+}$ and $p_{-}$ increase
with the threshold $\zeta$. Especially, $p_{+}$ and $p_{-}$ of every
$\zeta$ are equal within statistical errors. In other words, the
dynamic behavior at the microscopic time scale, typically in
minutes, is symmetric before and after large volatilities. However,
the exponents $p_{\pm}$ of the German DAX are somewhat larger than those of
the Chinese Indices.

\begin{figure}[p]\centering
\epsfysize=7.cm
\epsfclipoff
\fboxsep=0pt
\setlength{\unitlength}{0.7cm}
\begin{picture}(9,18)(0,0)
\put(-2,12.50){{\epsffile{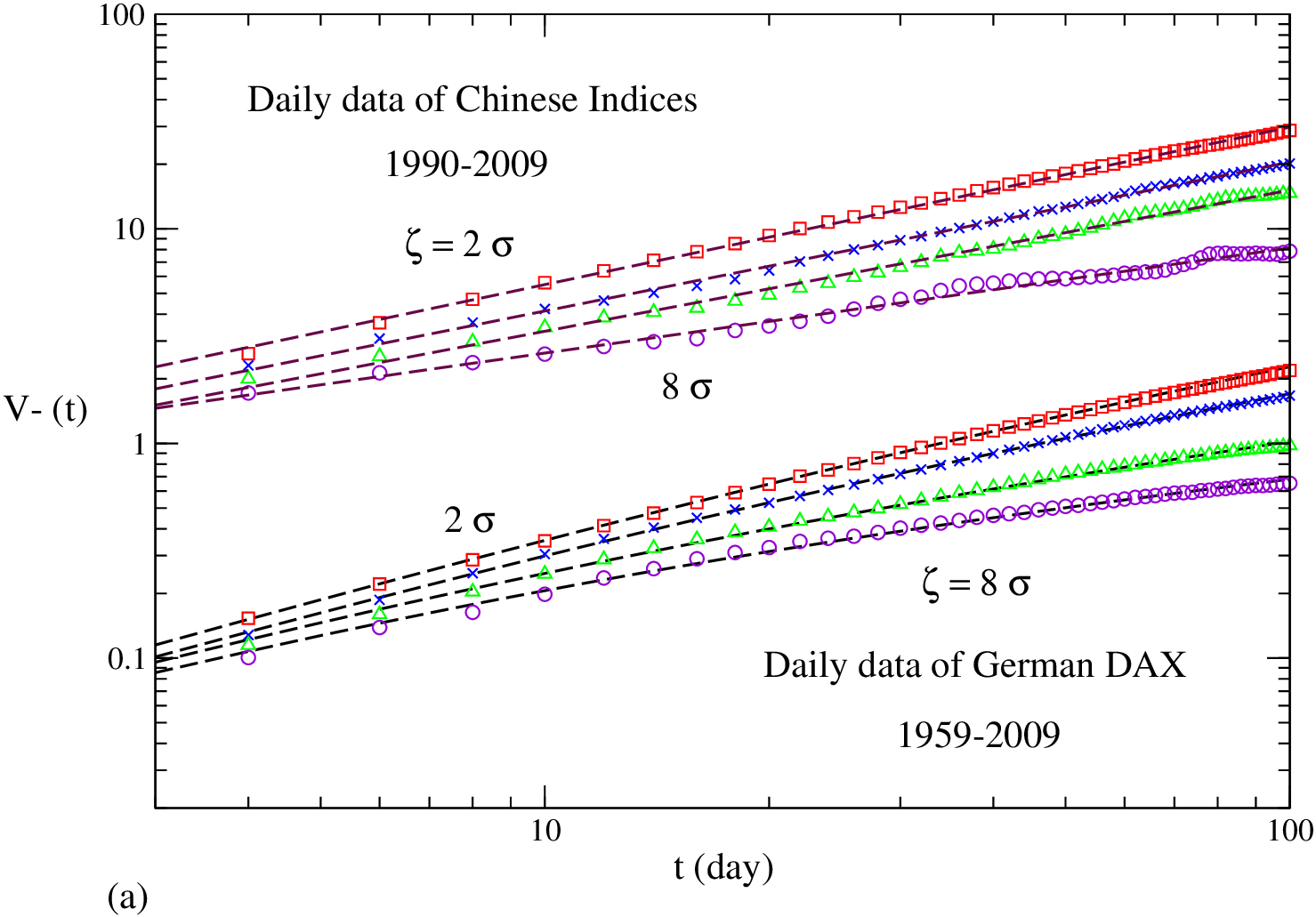}}}
\put(-2,0.50){{\epsffile{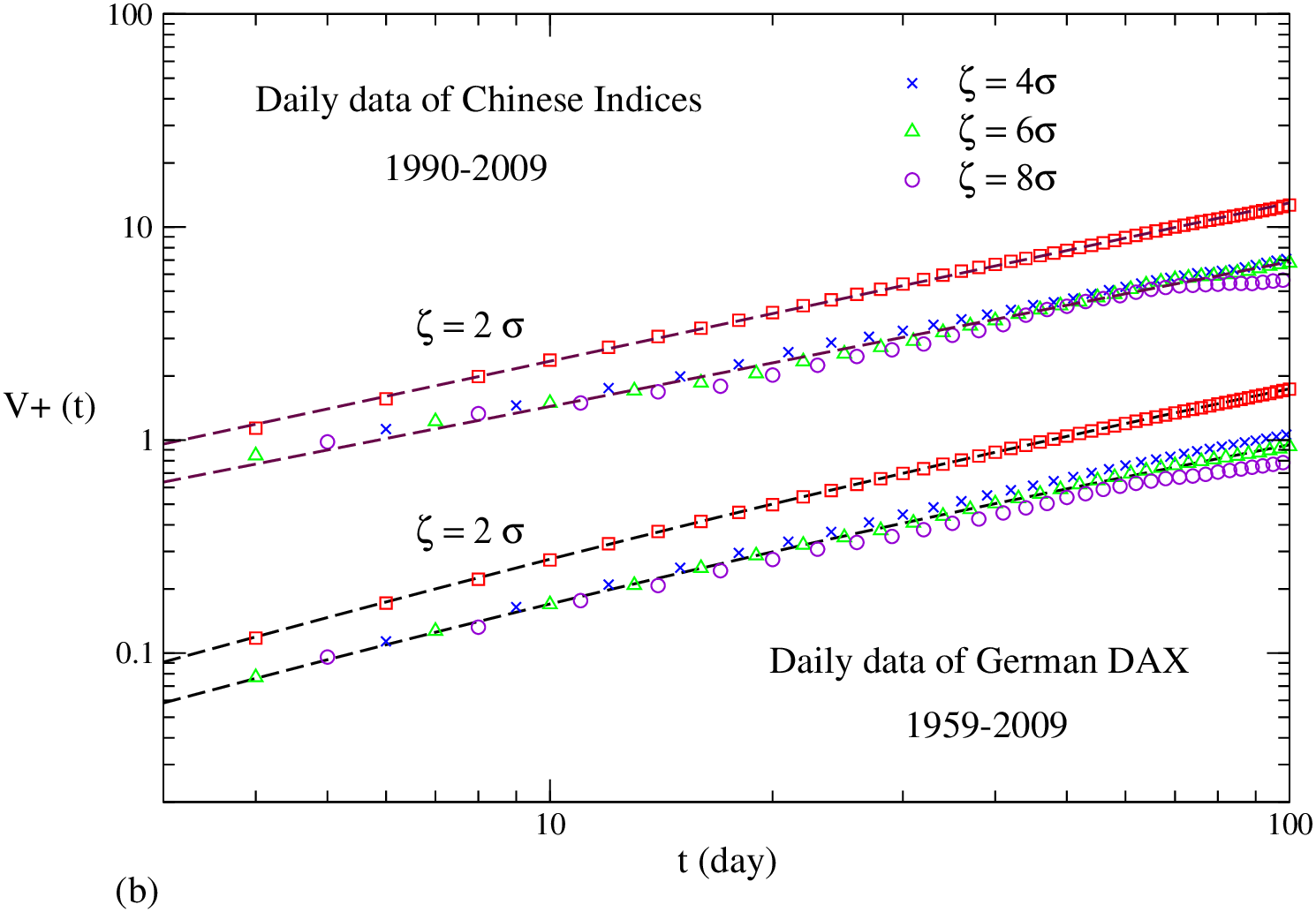}}}
\end{picture}
\caption{$V_\pm(t)$ for the daily data of the Chinese Indices and
German DAX. From above, the threshold is $\zeta= 2\sigma$,
$4\sigma$, $6\sigma$ and $8\sigma$ respectively. Dashed lines show
the power-law fits with Eq.~(\ref {e30}). $\tau_\pm=0$ for the
Chinese Indices and $\tau_\pm\neq 0$ for the German DAX. In (a), $p_-$
varies with $\zeta$. In (b), the $\zeta$-dependence of $p_+$ is weak. For
clarity, some curves have been shifted downwards or upwards.
}
\label{f2}
\end{figure}

To further understand the large-volatility dynamics of the financial
markets, we have also calculated $V_{\pm}(t)$ with the daily data of
the Chinese Indices and German DAX. As shown in Fig. \ref {f2}, the
dynamic behavior of $V_{\pm}(t)$ can also be described by Eq. (\ref
{e30}), although the curves look slightly fluctuating, compared with
those of the minute-to-minute data. From the fitting, we obtain $\tau_\pm
\approx 0$ for the Chinese Indices, while $\tau_\pm \neq 0$ for the
German DAX. The exponents $p_{\pm}$ are listed in the third and fourth sectors of Table~\ref{t1}.
$p_{\pm}$ of the daily data also
vary with $\zeta$, similar to those of the minute-to-minute data. However,
the $\zeta$-dependence of $p_{+}$ becomes obviously weaker, i.e.,
$p_{-}\neq p_{+}$. In other words, {\it the time-reversal symmetry before
and after large volatilities is violated at the daily time scale},
for both the Chinese Indices and German DAX.
Again $p_{\pm}$ of the German DAX are larger than those of the
Chinese Indices. But large part of the difference looks effectively induced by the short-time behavior.
If one directly estimates $p_{\pm}$ from the slopes of the tails of the curves,
$p_{\pm}$ of the German DAX are less different from those of the Chinese indices.
This is shown in the fifth sector of Table~\ref{t1}.
In fact, as a emerging market, the Chinese market
shares common features with the western markets in basic statistical
properties \cite{qiu07,qiu10}, while exhibits its own characteristics
in the return-volatility correlation and spatial structure
\cite{she09,she09a}.

In Refs. \cite{lil03,lil04}, the dynamic relaxation after a
financial crash, which is an event corresponding to an extremely
large $\zeta$ and with $R(t')<0$ in our terminology, has been
investigated. The observable $N_{+}(t)$, which is the number of
times that the volatility exceeds a certain threshold $\zeta_1$ in
the time $t$ after the financial crash, decays by a power law. For
comparison, we have also performed such an analysis. To reduce the
fluctuations, we choose a large but not extremely large threshold
$\zeta=12\sigma$ to gain some samples for average. Additionally we
extend the calculations to both $N_{+}(t)$ and $N_{-}(t)$. For the
minute-to-minute data, we observe that $N_{\pm}(t)$ of $\zeta_1=2\sigma$ to
$5\sigma$ could be fitted by Eq.~(\ref {e30}). The exponents
$p_{\pm}$ are weakly $\zeta_1$-dependent, and with $p_-=p_+$, i.e.,
similar to those for $V_\pm(t)$ in Table~\ref{t1}. For the daily
data, the fluctuations are large, although the asymmetric behavior
between $N_{+}(t)$ and $N_{-}(t)$ could be qualitatively observed.
The weak point of this analysis is that there are two thresholds
$\zeta$ and $\zeta_1$.

Up to here, we always average over all the selected large
volatilities in computing $V_{\pm}(t)$ (or $N_{\pm}(t)$). However,
the large volatilities could originate differently, and the dynamic
relaxation may depend on the category of the large volatilities.
Especially, it is puzzling {\it how the time-reversal asymmetry
arises at the daily time scale?} Our first thought is to classify
the large volatilities $|R(t')|$ by $R(t')<0$ and $R(t')>0$, i.e.,
the so-called "crashes" and "rallies". Such a classification is
not illustrating at all at the time scale in minutes, and the exponents $p_{\pm}$
remain the same for both the crashes and rallies.
For the daily data of the Chinese indices, $p_{\pm}$ of the crashes and rallies are also only marginally different.
The situation is subtle for the German DAX, and
will be clarified below.

\begin{figure}[p]\centering
\epsfysize=7.cm
\epsfclipoff
\fboxsep=0pt
\setlength{\unitlength}{0.7cm}
\begin{picture}(9,18)(0,0)
\put(-2,12.50){{\epsffile{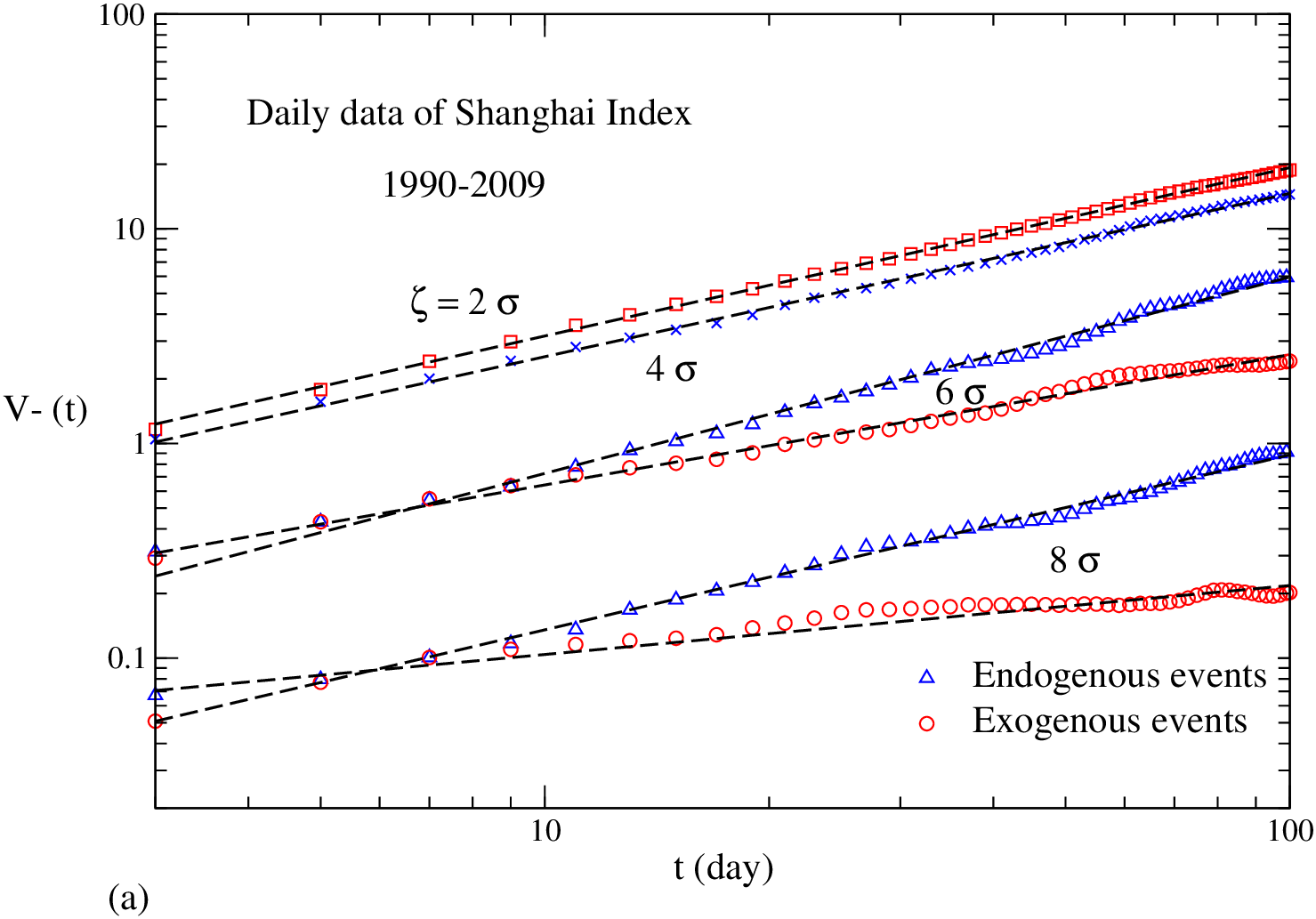}}}
\put(-2,0.50){{\epsffile{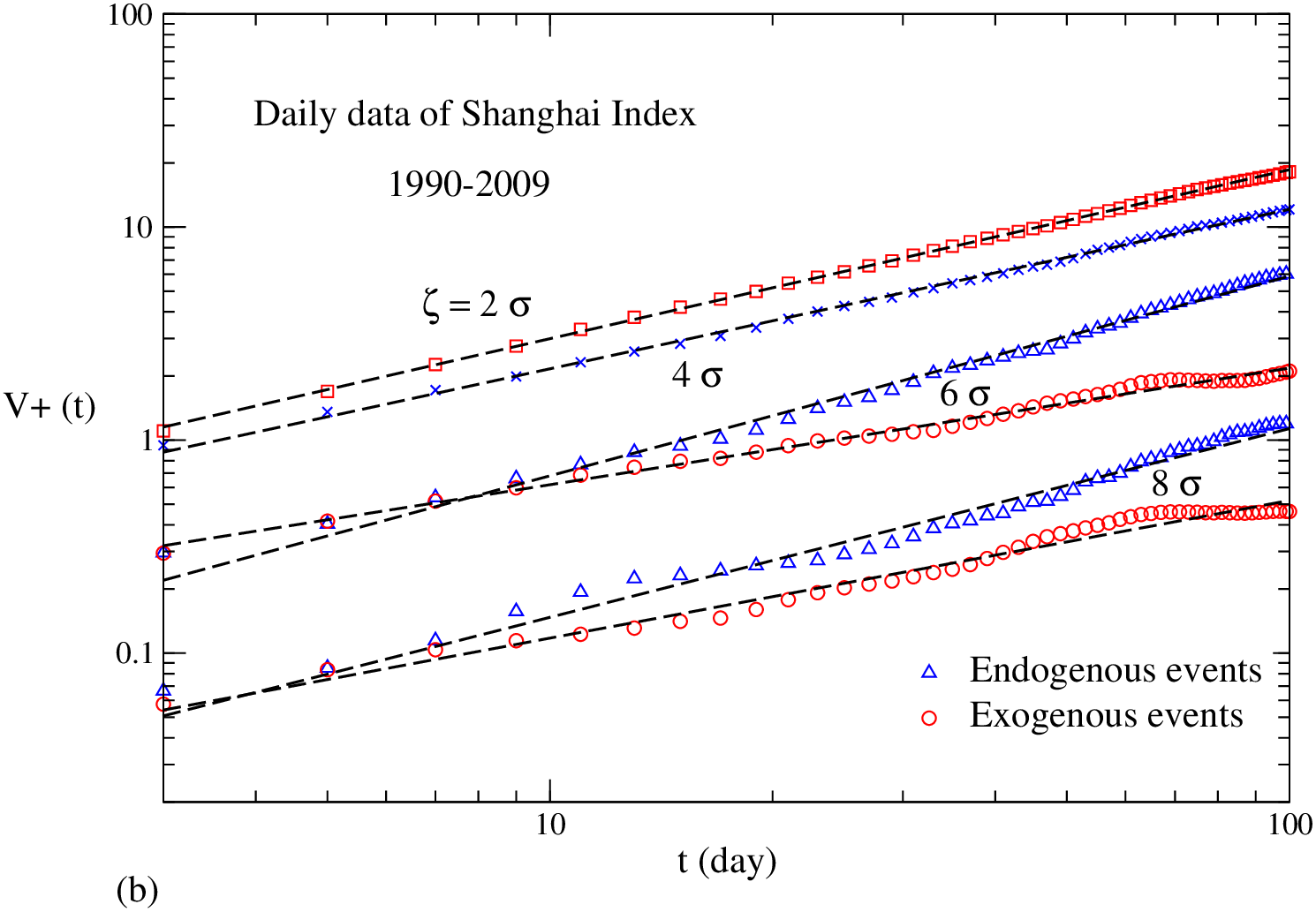}}}
\end{picture}
\caption{$V_\pm(t)$ for the daily data of the Shanghai Index. For
$\zeta= 6\sigma$ and $8\sigma$, $V_\pm(t)$ are displayed for
endogenous and exogenous events separately.
}
\label{f3}
\end{figure}

The large volatilities at the daily time scale could be also
classified into endogenous events and exogenous events
\cite{sor03,sor04}. An exogenous event is associated with the
market's response to external forces, and an endogenous event is
generated by the dynamic system itself. Naturally, different stock markets may differently respond
to the external forces. Looking carefully at the
history of the Shanghai stock market, for example, we find that
there are 9 exogenous events among the 16 large volatilities
selected by the threshold $\zeta=8\sigma$, as shown in Table \ref{t2}. For the large
volatilities corresponding to the thresholds such as $\zeta=2\sigma$
and $4\sigma$, it is not meaningful to naively identify the
external forces. In Fig.~\ref{f3}, $V_{\pm}(t)$ of $\zeta=6\sigma$
and $8\sigma$ are displayed for the endogenous and exogenous events
of the Shanghai Index. Obviously, the endogenous and exogenous
events lead to different dynamic behaviors for both $V_{+}(t)$ and $V_{-}(t)$. The dynamic relaxation
of the exogenous events is faster. For the Shenzhen Index, we obtain
qualitatively the same results but with somewhat larger
fluctuations.

\begin{figure}[p]\centering
\epsfysize=7.cm
\epsfclipoff
\fboxsep=0pt
\setlength{\unitlength}{0.7cm}
\begin{picture}(9,18)(0,0)
\put(-2,12.50){{\epsffile{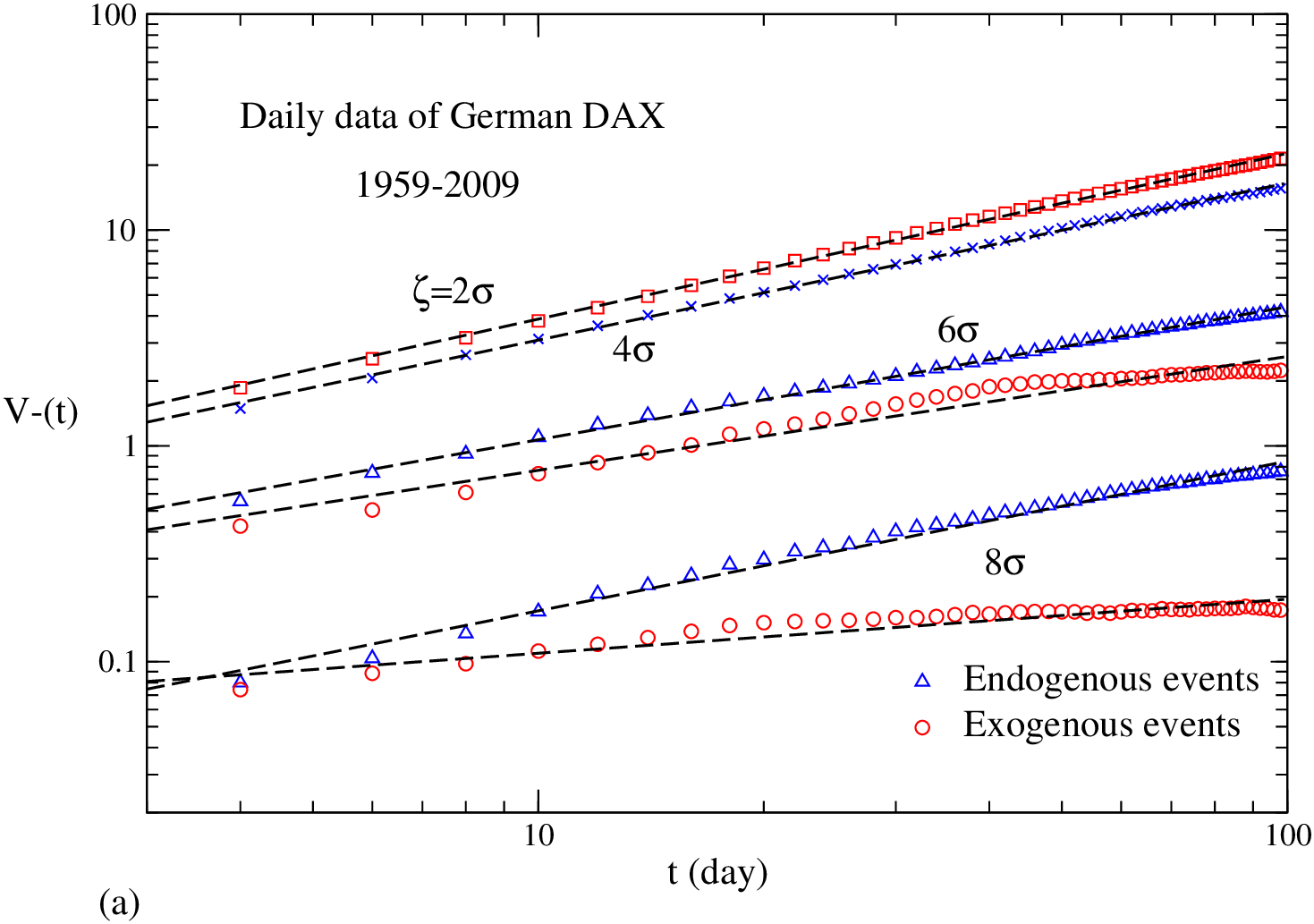}}}
\put(-2,0.50){{\epsffile{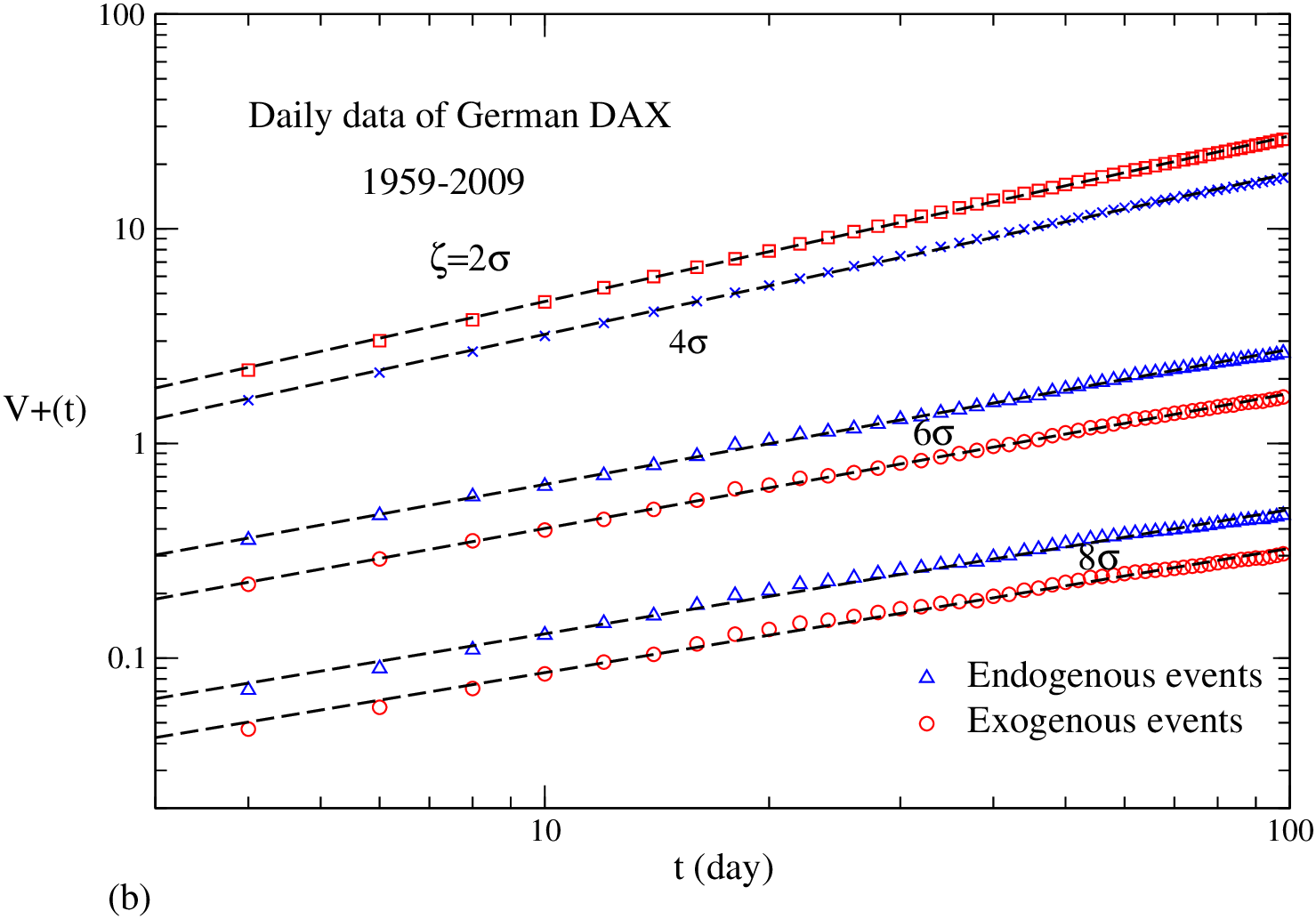}}}
\end{picture}
\caption{$V_\pm(t)$ for the daily data of the German DAX. For
$\zeta= 6\sigma$ and $8\sigma$, $V_\pm(t)$ are displayed for
endogenous and exogenous events separately.
}
\label{f4}
\end{figure}

For the German DAX, we have worked hard to identify the exogenous events,
in spite of our unfamiliarity of the history of the German stock market.
The findings are listed in Table \ref{t3}. In Fig.~\ref{f4}, $V_{\pm}(t)$ of $\zeta=6\sigma$
and $8\sigma$ are displayed for the endogenous and exogenous events
of the German DAX. Obviously, $V_{-}(t)$ exhibits different dynamic behaviors for
the endogenous and exogenous events, while $V_{+}(t)$ remains almost the same.
This behavior is different from that of the Shanghai Index. Although there exist some fluctuations,
we may estimate the exponents $p_{\pm}$ from the slopes of the curves, to quantify the findings in Figs.~\ref{f3} and \ref{f4}.
The results are given in Table~\ref{t4}.

Our first observation is that the time-reversal asymmetry in the large-volatility relaxation at the
daily time scale is mainly induced by the exogenous events, i.e., $p_{-}\approx p_{+}$ for the the endogenous events
and $p_{-}> p_{+}$ for the exogenous events. In particular, $p_{\pm}$ of the endogenous events are almost
independent of the threshold $\zeta$. For the German DAX, $p_{\pm}$ look increasing weakly with $\zeta$,
possibly because we might still miss some exogenous events.
It is due to the fluctuations that for the endogenous events with $\zeta=6\sigma$, $p_{-}=0.18$ and $p_{+}=0.16$ of the Shanghai Index
are somewhat smaller, and $p_{-}=0.37$ of the German DAX is slightly larger.

In financial markets, a large volatility does not necessarily
indicate that the dynamic system already jumps to a non-stationary
state, for the probability distribution of volatilities is with a
power-law tail. Since $p_{\pm}$ are
almost independent of $\zeta$ for the endogenous events, the dynamic system probably remains
still in the stationary state. However, the exogenous events may drive
the dynamic system to a non-stationary state, and thus induce the $\zeta$-dependent $p_{\pm}$ and
time-reversal asymmetry.

The second observation is that $p_{+}$ of the exogenous and endogenous events are almost equal for
the German DAX, while different for the Shanghai Index. Keeping in mind that $v_{-}(t)$ describes
how a financial market approaches a large volatility, it is understandable that
$p_{-}$ of the exogenous events is larger than that of the endogenous events,
since an exogenous event should emerge more radically than an endogenous event.
However, it is subtle whether $p_{+}$ of the exogenous and endogenous events are the same or different.
In principle, stock markets in different countries may differently respond to the external forces,
and different sorts of external forces may lead to different dynamic characteristics \cite{bre09,vru09}.
It seems that once exogenous or endogenous events occur, the German market does not distinguish
between exogenous and endogenous events. Such a phenomenon indicates that on the one hand,
that the German market is mature, and on the other hand, the exogenous events
in the Chinese and German markets may be different.
In fact, as shown in Tables \ref{t2} and \ref{t3}, the external forces could be grouped into two sorts:
market-policy changing, or political and economic accidents. All the exogenous events in the Shanghai
Index correspond to the market-policy changes, while those in the German DAX are induced by the political and economic accidents.
Since the Chinese market is newly set up, it is sensitive to the policies of the government.

The third observation is that all exogenous events except for one in the German DAX are crashes,
while those in the Shanghai Index are a mixture of crashes and rallies.
On the other hand, the endogenous events in the Shanghai Index are 4 rallies and 3 crashes,
while those in the German DAX are all rallies. In other words, the dynamic relaxation
of large volatilities is highly asymmetric between crashes and rallies in the German market,
different from that in the Chinese market. It should be also interesting and important that
for the mature German market, rallies are usually created by the endogenous events.

\section{Conclusion}
In summary, we have investigated the large-volatility dynamics in
financial markets, based on the minute-to-minute and daily data of the
Chinese Indices and German DAX. The dynamic relaxation before and
after large volatilities is characterized by the power law in
Eq.~(\ref{e30}). At the time scale in minutes, the exponents $p_\pm$
increase with the threshold $\zeta$, and the large-volatility
dynamics is time-reversal symmetric, i.e., $p_-=p_+$. At the daily
time scale, the $\zeta$-dependence of $p_+$ is weaker, and the
large-volatility dynamics is time-reversal asymmetric, i.e.,
$p_-\neq p_+$. Careful analysis reveals that not only the
time-reversal asymmetry but also the $\zeta$-dependence of $p_\pm$
are mainly induced by exogenous events. In this sense, the exogenous
events may drive the financial dynamics to a non-stationary state,
while the endogenous events may not.
Once exogenous or endogenous events occur, the German stock market does not distinguish
between exogenous and endogenous events, while the Chinese stock market does.
All the exogenous events in the Chinese market correspond to the market-policy changes,
while those in the German market are induced by the political and economic accidents.
It is interesting that
for the mature German market, rallies are usually created by the endogenous events.

\section*{Acknowledgements}
This work was supported in part by NNSF of China under Grant Nos. 10875102 and 11075137, and Zhejiang Provincial Natural
Science Foundation of China under Grant No. Z6090130.

\bibliographystyle{prsty}
\bibliography{eco2,zheng}

\begin{table}[p]\centering
 \small\begin{tabular}{c | c c c c }
\hline $\zeta$ &  $2\sigma$ &  $4\sigma$&  $6\sigma$
& $8\sigma$   \\
\hline
 &{CHN(min)} & & &\\
$p_{-}$      & 0.11(1)  & 0.15(1)  & 0.17(1) & 0.20(1) \\
$p_{+}$      & 0.11(1)  & 0.15(1)  & 0.18(1)  & 0.22(1)  \\
\hline & {DAX(min)}& & &\\
$p_{-}$      & 0.16(1)  & 0.23(1)  & 0.27(1) & 0.29(1) \\
$p_{+}$      & 0.16(1)  & 0.22(1)  & 0.26(1)  & 0.29(1)  \\
\hline\hline
& CHN(day) & & &\\
$p_{-}$      & 0.27(3) & 0.31(4) & 0.36(4) & 0.51(6)  \\
$p_{+}$      & 0.26(2) & 0.32(3) & 0.33(4) & 0.36(5)  \\
\hline & DAX(day)& & &\\
$\tau_{-}$   & 13.11    & 9.06     & 4.07     & 3.78     \\
$p_{-}$      & 0.41(3) & 0.47(4) & 0.60(5) & 0.77(7) \\
$\tau_{+}$   & 10.66    & 9.23     & 7.28     & 3.98     \\
$p_{+}$      & 0.40(2) & 0.42(3) & 0.45(5) & 0.46(5) \\
\hline
$p_-$ & $0.28(3)$ & $0.30(3)$ & $0.50(5)$ & $0.61(5)$\\
$p_+$ & $0.25(2)$ & $0.28(2)$ & $0.31(3)$ & $0.35(4)$\\
\hline
\end{tabular} \caption{$p_\pm$ are measured with the minute-to-minute
and daily data of the Chinese Indices (CHN) and German DAX. For CHN(min),
DAX(min) and CHN(day), $\tau_{\pm}=0$. In the fifth sector, $p_\pm$ are estimated
for DAX(day) from the tails of the curves in Fig. \ref{f2}.}\label{t1}
\end {table}

\begin{table}[h]\centering
\begin{tabular}{rcl}
\hline
Date &   Event
\tabularnewline\hline 92.05.21 & Rally & Chinese stock markets allowed free bidding transactions.
\tabularnewline94.03.14 & Rally &  The State Council announced that income from stock transfer \\
                        &       &    was exempt from tax this year, and banned the arbitrary right \\
                        &        &   issue of listed companies.
\tabularnewline94.08.01 & Rally &   CSRC decided to suspend new coming IPOs,  \\
                        &       &    to control the scale of right issues, and to develop mutual fund \\
                          &       &                         and fostered institution investors.
\tabularnewline95.05.18 & Rally &  CSRC suspended the pilot program of the national debt and \\
                        &       &   future trading.
\tabularnewline95.05.23 & Crash &  CSRC declared large amount of deposits of IPOs in 1995.
\tabularnewline96.12.16 & Crash &  CSRC set a limitation for the price change in a trading day \\
                        &       &   in stock markets.
\tabularnewline97.05.22 & Crash &  CSRC and Central Bank controlled the fund investing \\
                        &       &   in stock markets.
\tabularnewline01.10.23 & Rally &  CSRC declared stopping reduction of state-owned shares.
\tabularnewline08.09.19 & Rally &  MFSAT declared reduction of the stamp tax rate \\
                        &       &    in stock transactions; SASAC announced support for central \\
                        &       &    enterprises and for holdings of listed companies \\
                        &       &                           to buy shares back.
\tabularnewline
\hline
\end{tabular}
\caption{The $9$ exogenous events corresponding to $\zeta=8\sigma$ for the daily data of the Shanghai Index. The total number of the large volatilities for $\zeta=8\sigma$ is $16$.
IPOs: Initial Public Offerings; CSRC: China Security Regulatory Commission;
MFSAT: Ministry of Finance and State Administration of Taxation;
SASAC: State-owned Asset Supervision and Administration Commission.}
\label{t2}
\end{table}

\begin{table}[h]\centering
\begin{tabular}{rcl}
\hline
Date &   Event\tabularnewline
\hline
62.05.29 & Crash &  The big crash in NYSE in 05.28.\tabularnewline
87.10.19 & Crash &  Black Monday all over the world.\tabularnewline
87.10.22 & Crash &\tabularnewline
87.10.26 & Crash &\tabularnewline
87.10.28 & Crash &\tabularnewline
87.11.10 & Crash &\tabularnewline
89.10.16 & Crash &  Honecker in East Germany was forced to resign.\tabularnewline
91.01.17 & Rally &  Gulf War started.\tabularnewline
91.08.19 & Crash &  The coup against Gorbachev in Soviet Union.\tabularnewline
97.10.28 & Crash &  Asian Financial Crisis.\tabularnewline
01.09.12 & Crash &  Sep. 11 attack in US.\tabularnewline
08.01.21 & Crash &  Subprime mortgage crisis.\tabularnewline
08.10.06 & Crash & \tabularnewline
08.10.10 & Crash &  \tabularnewline
08.10.13 & Crash & \tabularnewline
08.11.06 & Crash & \tabularnewline
\hline
\end{tabular}
\caption{The $16$ exogenous events corresponding to $\zeta=8\sigma$ for the daily data of the German DAX.
The total number of the large volatilities for $\zeta=8\sigma$ is $27$.}
\label{t3}
\end{table}

\begin{table}[h]\centering
\centering \begin{tabular}{r|cccc|c}
\hline
$\zeta$ & $2\sigma$ & $4\sigma$ &$6\sigma$  & $8\sigma$ &\\
\hline
 & SHI(day) &  &  &   \\
$p_-$ & $0.22(2)$ & $0.24(2)$ &  $0.18(3)$ &   $0.20(6)$  & En. \\
     &           &           & $0.54(4)$ &  $0.80(7)$ & Ex.  \\
$p_+$ & $0.23(2)$ & $0.25(3)$ &  $0.16(4)$ &  $0.21(6)$ & En.  \\
     &           &           &  $0.52(4)$ &  $0.51(5)$ & Ex.  \\
\hline
 & DAX(day) &  &  &   &\\
$p_-$ & $0.28(3)$ & $0.30(3)$ & $0.37(3)$ &   $0.31(4)$ &En.  \\
     &           &           &$0.50(4)$  &  $0.76(6)$   & Ex.  \\
$p_+$ & $0.25(2)$ & $0.28(2)$ & $0.31(3)$ &  $0.34(5)$   &En. \\
     &           &           &$0.32(4)$  &  $0.35(4)$  & Ex.  \\
\hline
\end{tabular}
\caption{$p_\pm$ of the endogenous (EN.) and exogenous (EX.) events for the daily data of the Shanghai Index (SHI) and DAX.}
\label{t4}
\end{table}

\end{document}